\documentclass[prd,preprint,nofootinbib]{revtex4}

\usepackage{graphicx}
\usepackage{amssymb}
\usepackage{amsmath}
\usepackage{color}

\begin{document}

\renewcommand{\thefootnote}{\#\arabic{footnote}}
\newcommand{\rem}[1]{{\bf [#1]}}
\newcommand{\gsim}{ \mathop{}_ {\textstyle \sim}^{\textstyle >} }
\newcommand{\lsim}{ \mathop{}_ {\textstyle \sim}^{\textstyle <} }
\newcommand{\vev}[1]{ \left\langle {#1}  \right\rangle }
\newcommand{\bear}{\begin{array}}  
\newcommand {\eear}{\end{array}}
\newcommand{\bea}{\begin{eqnarray}}   
\newcommand{\eea}{\end{eqnarray}}
\newcommand{\beq}{\begin{equation}}   
\newcommand{\eeq}{\end{equation}}
\newcommand{\bef}{\begin{figure}}  
\newcommand {\eef}{\end{figure}}
\newcommand{\bec}{\begin{center}} 
\newcommand {\eec}{\end{center}}
\newcommand{\non}{\nonumber}  
\newcommand {\eqn}[1]{\beq {#1}\eeq}
\newcommand{\la}{\left\langle}  
\newcommand{\ra}{\right\rangle}
\newcommand{\ds}{\displaystyle}
\newcommand{\red}{\textcolor{red}}

\def\SEC#1{Sec.~\ref{#1}}
\def\FIG#1{Fig.~\ref{#1}}
\def\EQ#1{Eq.~(\ref{#1})}
\def\EQS#1{Eqs.~(\ref{#1})}
\def\lrf#1#2{ \left(\frac{#1}{#2}\right)}
\def\lrfp#1#2#3{ \left(\frac{#1}{#2} \right)^{#3}}
\def\GEV#1{10^{#1}{\rm\,GeV}}
\def\MEV#1{10^{#1}{\rm\,MeV}}
\def\KEV#1{10^{#1}{\rm\,keV}}

\def\REF#1{(\ref{#1})}
\def\lrf#1#2{ \left(\frac{#1}{#2}\right)}
\def\lrfp#1#2#3{ \left(\frac{#1}{#2} \right)^{#3}}

\begin{flushright}
\end{flushright}

\title{
Cosmological Moduli Problem from Thermal Effects
}

\author{Kazunori Nakayama$^{1}$ and  Fuminobu Takahashi$^{2}$}

\affiliation{
${}^1$Institute for Cosmic Ray Research,
     University of Tokyo, Chiba 277-8582, Japan\\
${}^2$Institute for the Physics and Mathematics of the Universe,
University of Tokyo, Chiba 277-8568, Japan }

\date{\today}

\begin{abstract}
We estimate the cosmological abundance of a modulus field that has
dilatonic couplings to gauge fields, paying particular attention to
thermal corrections on the modulus potential. We find that a certain amount
of the modulus coherent oscillations is necessarily induced by a linear thermal effect. 
We argue that such an estimate provides the smallest possible 
modulus abundance for a given thermal history of the Universe.
As an example we apply our results to a saxion,   a bosonic supersymmetric partner of 
an axion, and derive a tight bound on the reheating temperature. 
We emphasize that the problem cannot be avoided by fine-tuning the initial deviation of the modulus
field,  since the minimal amount of the modulus is induced by the dynamics of the scalar potential.
\end{abstract}

\preprint{IPMU 08-0085}
\preprint{ICRR-Report-533}
\pacs{98.80.Cq}

\maketitle

\section{Introduction}
\label{sec:1}
Many scalar fields are expected to be present in Nature,
and some of them may play important roles in cosmology. In supergravity and
string theories, there are modulus fields, some of which remain light and acquire 
masses from supersymmetry (SUSY) breaking and non-perturbative effects. 
They have interactions with the standard-model (SM) particles typically suppressed 
by a high energy scale such as the grand unification theory (GUT) scale or the Planck scale,
and it is known that  those modulus fields induce a notorious
cosmological moduli problem~\cite{ModuliProblem}. Recently the problem turned out to be
much more acute than previously thought, since the modulus decay generically
produces too many gravitinos which will significantly affect the standard
cosmology~\cite{Endo:2006zj,Dine:2006ii}. 

After reheating of the inflation,  the universe will be filled with a hot thermal plasma of
the SM particles.
If the modulus field couples to the SM particles (or any particles
in thermal plasma), the potential gets generically modified, which may affect
the cosmological evolution of the modulus field. In fact, it was pointed out 
in Refs.~\cite{Dine:2000ds,Buchmuller:2003is,Buchmuller:2004xr} 
that the modulus potential can be significantly modified
especially if the modulus possesses a dilatonic coupling, which induces a correction that is
linear in the modulus field. In particular, such a thermal effect may destabilize the
modulus after inflation, setting a tight bound on the highest temperature of the universe
\cite{Buchmuller:2004xr}.

In this paper we rigorously estimate the modulus abundance when such
a linear thermal correction is present, and find that the modulus abundance is bounded below by
a non-vanishing value. Although it was known to some people that the linear thermal
correction induces a certain amount of the modulus~\cite{Buchmuller:2003is}, it has not been studied how serious
the resultant cosmological moduli problem would be. Since such an estimate
provides an absolute minimum of the modulus abundance for a given thermal history of
the Universe, our discussion is conservative and generic. In particular, one cannot avoid 
the problem by tuning the initial deviation of the modulus field (e.g. based on an anthropic argument).
As an example we will apply our result to the saxion \cite{Rajagopal:1990yx}, 
a bosonic supersymmetric partner of an axion \cite{Peccei:1977hh}, 
and derive a tight upper bound on the reheating temperature. For the Peccei-Quinn (PQ)
scale $f_a = {\cal O}(10^{16})$\,GeV and the saxion mass $m_{s} = {\cal O}(10)$\,eV,
we will see that the reheating temperature should be lower than $10^3$~GeV, which is
in conflict with the thermal leptogenesis scenario \cite{Fukugita:1986hr}.

\section{Thermal effects on the modulus potential and minimum modulus abundance}
\label{sec:2}

As pointed out in Refs.~\cite{Dine:2000ds,Buchmuller:2003is,Buchmuller:2004xr}, 
a modulus potential receives corrections from thermal effects. 
The free energy of the SUSY $SU(N_c)$ QCD
with $N_f$ flavors in the fundamental representation,
assuming that the temperature is high enough to thermalize all these species, reads
\begin{equation}
	\mathcal F (T,\phi) = -\frac{\pi^2 T^4}{24}
	\left[ a - b g_s(\phi)^2 + \mathcal O (g_s(\phi)^3)\right],  \label{FT}
\end{equation}
where $a = 2N_c N_f + N_c^2-1$ and $b= (3/8\pi^2)(N_c^2-1)(N_c+3N_f)$, and
$g_s$ denotes the QCD gauge coupling constant, 
which is in general depends on the modulus field value ($\phi$).
This yields finite-temperature effective potential for the modulus.
In particular, we focus on the the following linear term,
\beq
	V_T(\phi) = - \kappa \frac{T^4}{M_P} \phi,   \label{linear}
\eeq
where the coefficient $\kappa$ depends on the model.
Since the modulus potential is time-dependent, some amount of the modulus
condensate will be necessarily produced, which is the main concern
of this letter.
Possible effects of the thermal mass term will be discussed later.



As a toy model, we consider a modulus field whose potential is given by\footnote{
	Although there may be a Hubble mass term $\sim c^2H^2 (\phi-\phi_0)^2$ with
	some arbitrary value of $\phi_0$, inclusion of this term
	does not modify the following argument unless the coefficient $c$ is much larger than one
	\cite{Linde:1996cx}.
}
\begin{equation}
	V(\phi) = \frac{1}{2}m^2 \phi^2 - \kappa \frac{T^4}{M_P}\phi.
\end{equation}
In the absence of the temperature-dependent linear term, one can tune the initial position of the
modulus field as $\phi \sim 0$ to suppress the modulus abundance,
possibly based on anthropic arguments.
However, the linear term shifts the position of the potential minimum
in a time-dependent way,
and this dynamically induces a coherent motion of the modulus field.
The typical amplitude of the motion induced by this effect is of the order 
$\delta \phi \sim \kappa T^4/(m^2 M_P)$, and hence 
the minimum modulus abundance is estimated as
\begin{equation}
	\frac{\rho_\phi}{s}\sim \left \{
	\begin{array}{ll}
	\displaystyle \frac{45\kappa^2}{4\pi^2 g_*} \frac{T_{\rm os}^5}{m^2M_P^2} 
	~~~{\rm for}~~~T_{\rm os} < T_{\rm R} \\
	\displaystyle \frac{45\kappa^2}{4\pi^2 g_*} \frac{T_{\rm R}^5}{m^2M_P^2}
	~~~{\rm for}~~~T_{\rm os} > T_{\rm R}
	\end{array} \right. , \label{min}
\end{equation}
where $T_{\rm os}$ is the temperature at which the modulus begins to oscillate,
and $T_{\rm R}$ is the reheating temperature after inflation.
This can be evaluated as
\begin{equation}
	\frac{\rho_\phi}{s}\sim \left \{
	\begin{array}{ll}
	\displaystyle 1.4\times 10^6~{\rm GeV} \kappa^2
	\left ( \frac{g_*}{228.75} \right )^{-9/4}
	\left ( \frac{m_\phi}{100~{\rm GeV}} \right )^{1/2}
	&~~~{\rm for}~~~T_{\rm os} < T_{\rm R} \\
	\displaystyle 8.4\times 10^{-4}~{\rm GeV} \kappa^2
	\left ( \frac{g_*}{228.75} \right )^{-1}
	\left ( \frac{m_\phi}{100~{\rm GeV}} \right )^{-2}
	\left ( \frac{T_{\rm R}}{10^9~{\rm GeV}} \right )^{5}
	&~~~{\rm for}~~~T_{\rm os} > T_{\rm R}  \label{min2}
	\end{array} \right. .
\end{equation}
Here we have assumed that the temperature of the dilute plasma before the reheating completes
is given by $T^4 \sim T_{\rm R}^2 HM_P$, where $H$ is the Hubble parameter \cite{KT}.
Note that in order for the above analysis to be valid, $\delta \phi$ must be much smaller than 
$\kappa^{-1}M_P$,
and this sets an upper bound on the temperature as $T < T_c\sim (mM_P/\kappa)^{1/2}$.
For $\kappa \lsim 1$, the critical temperature $T_c$ is always higher than $T_{\rm os}$, and so,
the above estimate on the modulus abundance is valid. It is model-dependent what the dynamics
of the modulus would be at an temperature above $T_c$. For instance, the modulus potential
may be destabilized and the modulus field may start rolling toward infinity~\cite{Buchmuller:2004xr}.
Since our concern here is the modulus dynamics at a temperature below $T_{\rm os}$,
the minimal modulus abundance (\ref{min2}) is a conservative one.
We have numerically checked that Eq.~(\ref{min}) provides the minimum modulus abundance,
which cannot be reduced further by tuning the initial value of the modulus field.\footnote{
If we allow ourselves to choose an arbitrary initial velocity $\dot{\phi}$ as well, the modulus abundance can be
significantly reduced. However, the $\dot{\phi}$ is normally dependent on the potential and
the initial position, and therefore the modulus dynamics would not allow us to
do so. We thank K. Hamaguchi for useful comments on this issue.
}
As we will see, even this minimum abundance causes a cosmological disaster in general.

\section{Examples}
\label{sec:4}

\begin{figure}[tbp]
 \begin{center}
   \includegraphics[width=0.8\linewidth]{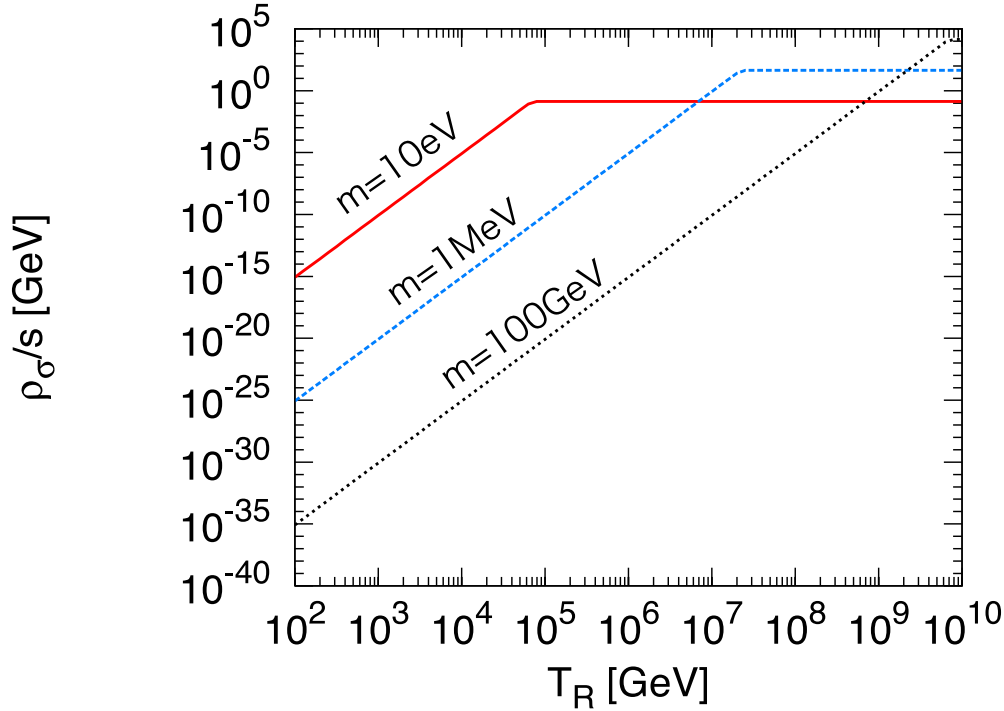}
   \caption{The minimum saxion abundances for $m_\sigma=10$~eV, $1$~MeV, and $100$~GeV
   as a function of the reheating temperature. Here we take $\kappa = 0.1$,
   corresponding to $f_a \sim 5\times 10^{17}~$GeV.}
   \label{fig1}
 \end{center}
\end{figure}

We have seen that there exists a strict ``minimum'' abundance of moduli,
dynamically induced by the thermal effects.
In order to see that such an effect makes cosmological moduli problems worse
than previously thought,
let us consider the SUSY axion model \cite{Rajagopal:1990yx}.
We denote an axion supermultiplet by $\mathcal A$
and the PQ scale by $f_a$.
Here we assume $f_a \gtrsim 10^{16}~$GeV, motivated by string axion models \cite{Svrcek:2006yi}.
The interaction of the axion multiplet $\mathcal A$ to the $SU(3)_C$ gluon
gauge multiplet is given by
\beq
{\cal L} \;=\; \int d^2 \theta \frac{\mathcal A}{32 \pi^2 f_a} W^\alpha W_\alpha +{\rm h.c.}.
\eeq
The imaginary lowest component of $\mathcal A$ is
the axion $a$  which is to solve the strong CP problem, whereas the
real component is the saxion $\sigma$:
\beq
\mathcal A \;\equiv\; \frac{\sigma+ia}{\sqrt{2}}.
\eeq
Let us consider the thermal effects on the saxion potential.
The QCD gauge coupling constant $g_s$ in Eq.~(\ref{FT}) depends on $\sigma$ as
\begin{equation}
	\frac{1}{g_s(\sigma)^2} = \frac{1}{g_s(0)^2} + \frac{\sigma}{8\pi^2 f_a}.
\end{equation}
Thus this leads to the finite-temperature effective potential of the form (\ref{linear}) for the saxion.
Note that the second term must be regarded as a small correction in order for the analysis to be valid.
Substituting $N_c=3$ and $N_f=6$ in the SUSY SM, $\kappa$ in Eq.~(\ref{linear}) is given by\footnote{
The value of $\kappa$ is of order unity for $f_a \gsim 10^{16}$\,GeV, and so, $T_{\rm os}$ 
can be comparable to $T_c$ if the reheating is completed before the commencement of
the oscillations, i.e., $T_{\rm R} > T_c$. On the other hand, we are more interested in a case of 
$T_{\rm R} < T_{\rm os}$,
where $T_{\rm os} \ll T_c$ is satisfied and therefore our analysis is valid.
}
\begin{equation}
	\kappa = \frac{21}{64\pi^2}g_s^4\left ( \frac{M_P}{f_a} \right )
	\sim 0.02\left(\frac{M_P}{f_a}\right).
\end{equation}
Generally, the saxion has a mass of order of the gravitino and 
it is known that the coherent oscillations of the saxion cause cosmological problems
for a wide range of the saxion mass \cite{Banks:1996ea,Asaka:1998xa,Kawasaki:2007mk}.
Actually there exists a minimum abundance of the saxion induced by thermal effect
given by Eq.~(\ref{min}), and even such a minimum abundance of the saxion
has significant impacts on cosmology.
In Fig.~\ref{fig1} we show the minimum saxion abundance for $m_\sigma=10$~eV, 1~MeV, 100~GeV
as a function of the reheating temperature $T_{\rm R}$.
For $m_\sigma=10$~eV, the constraint on the saxion abundance comes from the requirement that
the saxion must not exceed the dark matter abundance,
which restricts the saxion abundance as $\rho_\sigma/s \lesssim 4\times 10^{-10}$~GeV.
This translates into the bound on the reheating temperature, $T_{\rm R}\lesssim 10^3$~GeV.
Thus thermal leptogenesis scenario \cite{Fukugita:1986hr}
is incompatible with the cosmological saxion problem
for the ultra-light gravitino $m_{3/2} \sim 10~$eV.
For the intermediate mass scale $m_\sigma\sim \mathcal O$(keV)-$\mathcal O$(TeV),
the diffuse X($\gamma$)-ray background or big-bang nucleosynthesis (BBN) sets strong bounds
on the saxion abundance, 
and the reheating temperature cannot be as large as $T_{\rm R} \sim 10^9$~GeV.
On the other hand, if the saxion mass is heavy enough to decay before BBN,
the constraint can be evaded.
Thus we conclude that the thermal leptogenesis scenario is excluded in the SUSY axion model
for $f_a \gtrsim 10^{16}~$GeV except for the heavy gravitino (saxion) case, 
as is realized in the anomaly-mediated SUSY breaking models \cite{Randall:1998uk}.\footnote{
	The abundance of the axion ($a$) can be negligibly small by tuning the initial position
	of the axion.
	However, the axion always has isocurvature fluctuation \cite{Seckel:1985tj}
	as well as possibly large non-Gaussianity \cite{Linde:1996gt,Boubekeur:2005fj,Kawasaki:2008sn}
	and an upper bound on the inflation scale is imposed in order to be consistent with 
	cosmological observations.
}



On the other hand, if the PQ scale $f_a$ is in the ordinary axion window 
$10^{10}\,{\rm GeV} \lsim f_a \lsim 10^{12}\,$GeV, and if the reheating temperature is high enough
so that the thermal leptogenesis works, the saxion may be in thermal equilibrium.
In this case the cosmological saxion abundance
will be in conflict with either BBN or dark matter abundance
except for $m \lsim {\cal O}(10)$\,eV or $m\gtrsim {\mathcal O}(10)$~TeV.


\section{Discussion and Conclusions}
\label{sec:5}


We have discussed that thermal effects generically produce a linear term in the 
modulus potential, which dynamically induces a coherent motion of the modulus field.
This provides the smallest possible modulus abundance among all the possible initial 
positions of the modulus field.
Thus the cosmological constraints given in this letter is conservative.
The same analysis applies to any moduli which have dilatonic coupling with gauge fields.

We comment on the effect of a thermal mass term such as $\sim \lambda (T^4/f_a^2)\phi^2$,
where $\lambda \ll 1$ represents the one-loop factor as well as the gauge couplings.
We have neglected the term assuming that the thermal mass is smaller than the Hubble parameter
and therefore it does not affect the modulus dynamics for large $f_a (\gtrsim 10^{16}~{\rm GeV})$.
If the thermal mass term is effective, the modulus may be settled at the 
potential minimum due to the thermal mass.\footnote{
The modulus dynamics is model-dependent when the thermal mass plays an important role.
} In this case, the modulus may adiabatically follow the temporal minimum 
during the subsequent cosmological evolution
and the coherent oscillations of the modulus might  not  be induced
as noted in Ref.~\cite{Linde:1996cx} in the context of a large Hubble mass term,
although thermal production of saxion as well as axino~\cite{Brandenburg:2004du} 
may occur at an non-negligible rate.

Finally we comment on possible ways to relax the cosmological moduli problem.
As already noted, if the moduli are heavy enough,
the minimum abundance provided by Eq.~(\ref{min}) significantly decreases and
also they decay well before BBN.
For example, in the dynamical SUSY breaking models, the SUSY breaking field has 
mass of the order of the dynamical scale $\Lambda$, which is much larger than the gravitino mass.
Thus the upper bound on the reheating temperature is not so stringent in such a case.
Also if there were additional entropy production processes in the early Universe
\cite{Yamamoto:1985rd,Kawasaki:2004rx},
the moduli can be sufficiently diluted.
However, one should note that the pre-existing baryon asymmetry is also diluted
and only a few examples of baryogenesis mechanism are known to work
\cite{Dimopoulos:1987rk,Stewart:1996ai,Kasuya:2001tp}.

\begin{acknowledgements} 
This work was supported by World Premier
International Research Center Initiative (WPI Initiative), MEXT, Japan.
K.N. would like to thank the Japan Society for the Promotion of
Science for financial support.
\end{acknowledgements}




\begin{thebibliography}{10}
\bibitem{ModuliProblem}
%
  G.~D.~Coughlan et al
  Phys.\ Lett.\ B {\bf 131} (1983) 59;
  T.~Banks, D.~B.~Kaplan and A.~E.~Nelson,
  Phys.\ Rev.\ D {\bf 49} (1994) 779;
  B.~de Carlos et al
  Phys.\ Lett.\ B {\bf 318} (1993) 447.


\bibitem{Endo:2006zj}
  M.~Endo, K.~Hamaguchi and F.~Takahashi,
  Phys.\ Rev.\ Lett.\  {\bf 96}, 211301 (2006)
  [arXiv:hep-ph/0602061];
  S.~Nakamura and M.~Yamaguchi,
  Phys.\ Lett.\  B {\bf 638}, 389 (2006)
  [arXiv:hep-ph/0602081].
  
  
\bibitem{Dine:2006ii}
  M.~Dine, R.~Kitano, A.~Morisse and Y.~Shirman,
  Phys.\ Rev.\  D {\bf 73}, 123518 (2006)
  [arXiv:hep-ph/0604140].
  M.~Endo, K.~Hamaguchi and F.~Takahashi,
  Phys.\ Rev.\  D {\bf 74}, 023531 (2006)
  [arXiv:hep-ph/0605091].


\bibitem{Dine:2000ds}
  M.~Dine,
  Phys.\ Lett.\  B {\bf 482}, 213 (2000)
  [arXiv:hep-th/0002047].
  
  
\bibitem{Buchmuller:2003is}
  W.~Buchmuller, K.~Hamaguchi and M.~Ratz,
  Phys.\ Lett.\  B {\bf 574}, 156 (2003)
  [arXiv:hep-ph/0307181].


\bibitem{Buchmuller:2004xr}
  W.~Buchmuller, K.~Hamaguchi, O.~Lebedev and M.~Ratz,
  Nucl.\ Phys.\  B {\bf 699}, 292 (2004)
  [arXiv:hep-th/0404168].
  
  
\bibitem{Rajagopal:1990yx}
  K.~Rajagopal, M.~S.~Turner and F.~Wilczek,
  Nucl.\ Phys.\  B {\bf 358}, 447 (1991).
  
  
\bibitem{Fukugita:1986hr}
  M.~Fukugita and T.~Yanagida,
  Phys.\ Lett.\  B {\bf 174}, 45 (1986).

  
\bibitem{Peccei:1977hh}
  R.~D.~Peccei and H.~R.~Quinn,
  Phys.\ Rev.\ Lett.\  {\bf 38}, 1440 (1977).
  
  
\bibitem{Linde:1996cx}
  A.~D.~Linde,
  Phys.\ Rev.\  D {\bf 53}, 4129 (1996)
  [arXiv:hep-th/9601083].
  
\bibitem{KT}
  E.~Kolb and M.~Turner, {\it The Early Universe},   
  Addison-Wesley, CA, 1990.
  
  
\bibitem{Svrcek:2006yi}
  P.~Svrcek and E.~Witten,
  JHEP {\bf 0606}, 051 (2006)
  [arXiv:hep-th/0605206].
  
    
\bibitem{Banks:1996ea}
  T.~Banks and M.~Dine,
  Nucl.\ Phys.\  B {\bf 505}, 445 (1997)
  [arXiv:hep-th/9608197];
  T.~Banks, M.~Dine and M.~Graesser,
  Phys.\ Rev.\  D {\bf 68}, 075011 (2003)
  [arXiv:hep-ph/0210256].
  
  
\bibitem{Asaka:1998xa}
  T.~Asaka and M.~Yamaguchi,
  Phys.\ Rev.\  D {\bf 59}, 125003 (1999)
  [arXiv:hep-ph/9811451].
  
  
\bibitem{Kawasaki:2007mk}
  M.~Kawasaki, K.~Nakayama and M.~Senami,
  JCAP {\bf 0803}, 009 (2008)
  [arXiv:0711.3083 [hep-ph]];
  M.~Kawasaki and K.~Nakayama,
  Phys.\ Rev.\  D {\bf 77}, 123524 (2008)
  [arXiv:0802.2487 [hep-ph]].
  
    
\bibitem{Randall:1998uk}
  L.~Randall and R.~Sundrum,
  Nucl.\ Phys.\  B {\bf 557}, 79 (1999)
  [arXiv:hep-th/9810155];
  G.~F.~Giudice, M.~A.~Luty, H.~Murayama and R.~Rattazzi,
  JHEP {\bf 9812}, 027 (1998)
  [arXiv:hep-ph/9810442].
  
  
\bibitem{Seckel:1985tj}
  D.~Seckel and M.~S.~Turner,
  Phys.\ Rev.\  D {\bf 32}, 3178 (1985);
  M.~S.~Turner and F.~Wilczek,
  Phys.\ Rev.\ Lett.\  {\bf 66}, 5 (1991);
  A.~D.~Linde,
  Phys.\ Lett.\  B {\bf 259}, 38 (1991).
  
  
\bibitem{Linde:1996gt}
  A.~D.~Linde and V.~F.~Mukhanov,
  Phys.\ Rev.\  D {\bf 56}, 535 (1997)
  [arXiv:astro-ph/9610219].
  
  
\bibitem{Boubekeur:2005fj}
  L.~Boubekeur and D.~H.~Lyth,
  Phys.\ Rev.\  D {\bf 73}, 021301 (2006)
  [arXiv:astro-ph/0504046].
  
  
\bibitem{Kawasaki:2008sn}
  M.~Kawasaki, K.~Nakayama, T.~Sekiguchi, T.~Suyama and F.~Takahashi,
  arXiv:0808.0009 [astro-ph];
  arXiv:0810.0208 [astro-ph].
  
  
\bibitem{Brandenburg:2004du}
  A.~Brandenburg and F.~D.~Steffen,
  JCAP {\bf 0408}, 008 (2004)
  [arXiv:hep-ph/0405158].
  
  
\bibitem{Yamamoto:1985rd}
  K.~Yamamoto,
  Phys.\ Lett.\  B {\bf 168}, 341 (1986);
  G.~Lazarides, C.~Panagiotakopoulos and Q.~Shafi,
  Phys.\ Rev.\ Lett.\  {\bf 56}, 557 (1986);
  D.~H.~Lyth and E.~D.~Stewart,
  Phys.\ Rev.\  D {\bf 53}, 1784 (1996)
  [arXiv:hep-ph/9510204].


\bibitem{Kawasaki:2004rx}
  M.~Kawasaki and F.~Takahashi,
  Phys.\ Lett.\  B {\bf 618}, 1 (2005)
  [arXiv:hep-ph/0410158].
  
  
\bibitem{Dimopoulos:1987rk}
  S.~Dimopoulos and L.~J.~Hall,
  Phys.\ Lett.\  B {\bf 196}, 135 (1987).
  
  
\bibitem{Stewart:1996ai}
  E.~D.~Stewart, M.~Kawasaki and T.~Yanagida,
  Phys.\ Rev.\  D {\bf 54}, 6032 (1996)
  [arXiv:hep-ph/9603324];
  D.~h.~Jeong, K.~Kadota, W.~I.~Park and E.~D.~Stewart,
  JHEP {\bf 0411}, 046 (2004)
  [arXiv:hep-ph/0406136];
  M.~Kawasaki and K.~Nakayama,
  Phys.\ Rev.\  D {\bf 74}, 123508 (2006)
  [arXiv:hep-ph/0608335];
  G.~N.~Felder, H.~Kim, W.~I.~Park and E.~D.~Stewart,
  JCAP {\bf 0706}, 005 (2007)
  [arXiv:hep-ph/0703275].


\bibitem{Kasuya:2001tp}
  S.~Kasuya, M.~Kawasaki and F.~Takahashi,
  Phys.\ Rev.\  D {\bf 65}, 063509 (2002)
  [arXiv:hep-ph/0108171];
  M.~Kawasaki and K.~Nakayama,
  Phys.\ Rev.\  D {\bf 76}, 043502 (2007)
  [arXiv:0705.0079 [hep-ph]].
  


\end{thebibliography}
\end{document}